\documentclass{iauc}
\usepackage{graphicx}
\pubyear{2005}
\volume{199}
\pagerange{1--}
\jname{Probing Galaxies through Quasar Absorption Lines}
\editors{P. R. Williams, C. Shu, and B. M\'{e}nard, eds.}

\def\Msun{M_\odot}

\title[Multi-Phase Galaxy Formation]
{Multi-Phase Galaxy Formation\\ and Quasar Absorption Systems}

\author[Ariyeh H. Maller]
{Ariyeh H. Maller}

\affiliation{Astronomy Department, University of Massachusetts Amherst, 
710 N. Pleasant St., Amherst, MA 01003, USA \break 
email: ari@astro.umass.edu\\}

\begin{document}

\maketitle

\begin{abstract}
The central problem of galaxy formation is understanding the cooling and 
condensation of gas in dark matter halos.  It is now clear that to match
observations this requires further physics than the simple assumptions 
of single phase gas cooling.  A model of multi-phase cooling 
\cite[(Maller \& Bullock 2004)]{mb:04}
can successfully account for the upper cutoff in the masses of galaxies
and provides a natural explanation of many types of absorption systems 
\cite[(Mo \& Miralda-Escude 1996)]{mm:96}.  

Absorption systems are our best probes of the gaseous content of galaxy 
halos and therefore provide important constraints on models for gas cooling
into galaxies.  All physical processes that effect gas cooling
redistribute gas and therefore are detectable in absorption systems. 
Detailed studies of the nature of gas in galaxy halos using absorption systems are
crucial for building a correct theory of galaxy formation.

\keywords{galaxies: mass function, galaxies: formation, quasars: absorption lines}

\end{abstract}

\firstsection 
\section{Multi-Phase Cooling}

\begin{figure}
 \includegraphics[width=\linewidth]{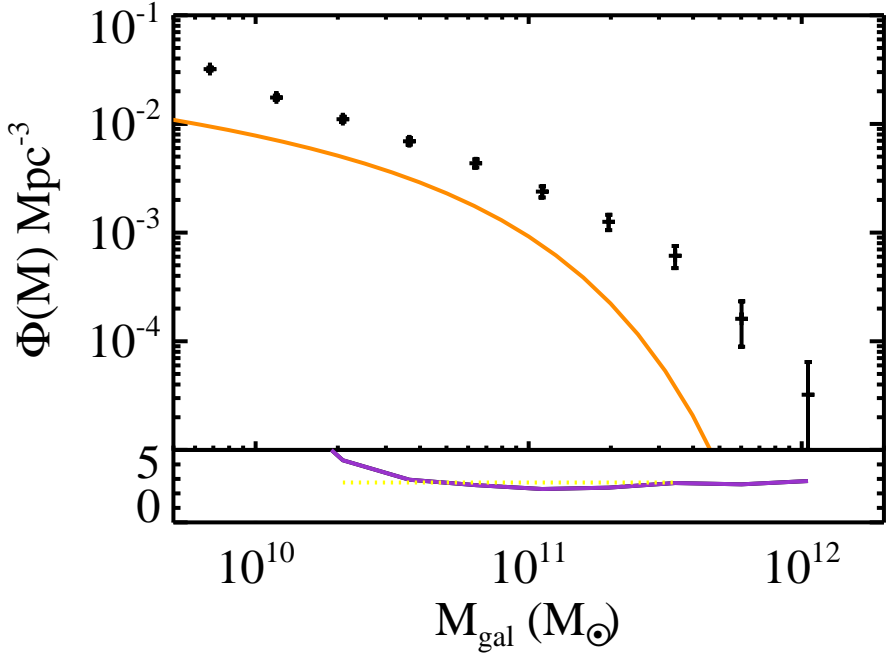}
  \caption{The upper panel shows the cumulative galaxy mass function at $z=0$ 
for a hydrodynamical simulation and, for comparison, the Schechter function fit of 
\cite[Bell et~al. (2003)]{bell:03}. The y-axis is in comoving units with $h=0.7$.    
The simulation produces far too many low and high mass 
galaxies, but galaxies around the bend in the Schechter function 
are only a factor of 2-3 too massive.  The bottom panel shows the factor 
by which the simulated galaxies masses should be divided in order to agree
with the observations. For the galaxy masses of 
$2 \times 10^{10} \Msun < M_{gal} < 6 \times 10^{11} \Msun$, 
the correction is roughly a factor of 2.75 as shown by the dotted line and 
does not depend strongly on galaxy mass.
}\label{fig:mf}
\end{figure}

The modern theory of galaxy formation rests on the concept of gas 
cooling in dark matter halos \cite{wr:78}.
However, simple models of single phase cooling predict a mass function
of galaxies significantly higher than observed.  
Figure \ref{fig:mf}, taken from \cite{maller:05},
shows the cumulative mass function of galaxies in an hydrodynamical
simulation compared to the observations of \cite{bell:03}.  The number
density of galaxies at a fixed mass is too high at all masses, but becomes
increasingly discrepant at high and low masses.  Thus the mass of cooled gas
must be significantly reduced from the simple single phase 
cooling implemented in the hydrodynamical simulation, especially in high and
low mass halos. 
 
\begin{figure}
 \includegraphics[width=\linewidth]{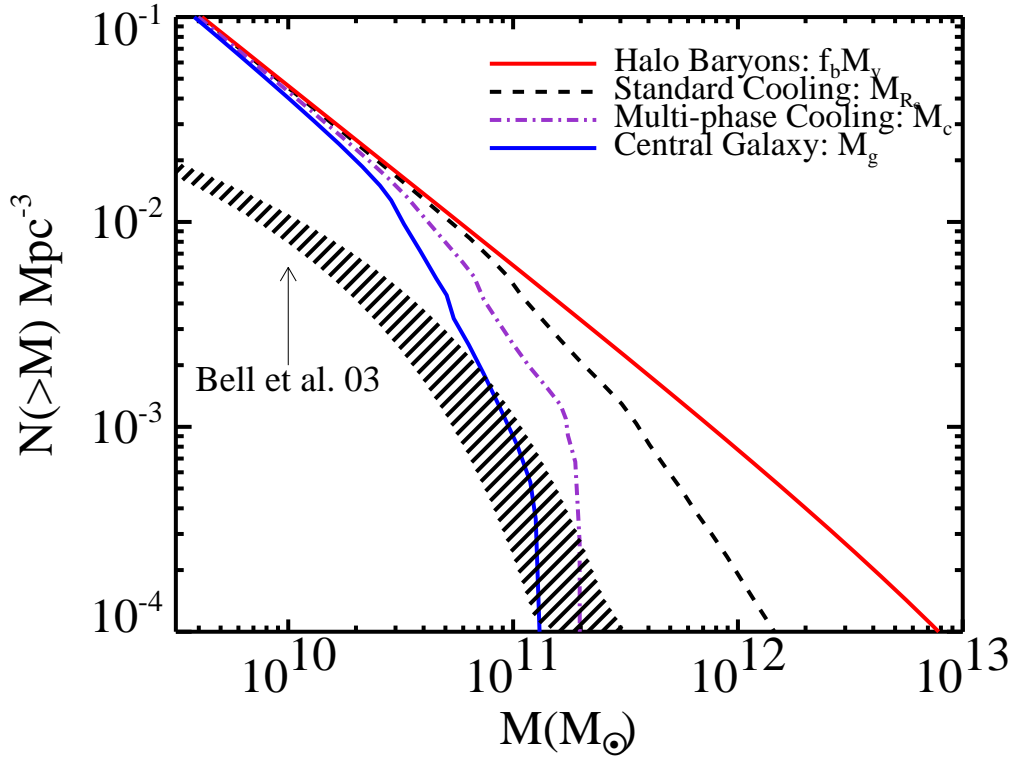}
 \caption{The cumulative baryonic mass function of galaxies reported by 
\protect{\cite[Bell et~al. (2003)]{bell:03}} (shaded band) compared to the cumulative mass 
function of halo baryons (top solid line). The short-dashed line shows the (central)
galaxy mass function that arises from assuming all of the mass within each halo's 
cooling radius cools onto the central galaxy, This is the same result seen in the
hydrodynamical simulations (previous Figure).  The dot-dashed line is the cooled mass 
function that arises in the \cite{mb:04} picture, which allows for the presence of a hot
corona in each halo.  Finally, the lowest solid line shows the central galaxy mass 
function that results from modeling the survival probability of cooled
clouds in the halo, assuming a typical cloud mass of $5 \times 10^6 \Msun$.
Only clouds that fall to the center of each halo are assumed to contribute to the 
central galaxy.  No merging has been accounted for in this estimate.  Merging will 
tend to populate the massive tail of the galaxy mass function, likely
bringing it even more closely in line with what is observed.
}\label{fig:mp}
\end{figure}

One solution to the problem for high mass halos is to consider multi-phase cooling.  
A cooling astrophysical plasma is thermally unstable and will naturally fragment and 
form a two-phase medium \cite{field:65}.  This means that a significant fraction 
of the halo gas can remain in a hot low-density galactic corona, while the remaining
gas cools into warm clouds and settles into a galactic disk only after loosing kinetic 
energy from collisions or ram pressure \cite[(Maller \& Bullock 2004)]{mb:04}.  
The result of such a model is to significantly reduce cooling in massive dark matter halos, 
bringing the high mass end of the galaxy mass function into agreement with observations 
(see Figure \ref{fig:mp}).  Specifically, for a galaxy in a Milky Way size halo if 
we take the warm clouds to have a mass of $10^6 \Msun$ then the mass of the galaxy's 
disk is half of what it would be in single-phase cooling and in good agreement with 
estimates of the Milky Way's mass.  For the gas that does not end up in the disk, 
two thirds is in a hot galactic corona while one third is in the form of warm clouds.  

Besides the agreement with the high mass end of the observed stellar mass function we 
would like to see observational evidence for the multi-phase nature of halo gas.  
\cite{mb:04} argue that the population of high velocity clouds seen in HI around the 
Milky Way represent this fragmented cloud population, the neutral component of mostly 
ionized $10^4 K$ clouds.  For other galaxies, quasar absorption systems are the best 
way to verify this picture.  Also, while we know that the cooling is thermally unstable, 
we do not know what the resulting distribution of warm cloud masses should be and how 
it might depend on halo mass and redshift.  Thus studying these clouds in differing 
environments is crucial to understanding the effect of multi-phase cooling on galaxy 
formation.

It has long been assumed that quasar absorption systems can be identified with the 
gaseous content of galaxy halos \cite{bs:69}. In fact the very multi-phase model 
discussed here has been proposed as the source of Lyman limit and metal-line systems
by \cite{mm:96}.  In the next section I describe the gaseous content of a galaxy's halo
and how observations of absorption systems may help us constrain their properties.

\section{The gaseous content of a galaxy's halo}

\begin{figure}
 \includegraphics[width=\linewidth]{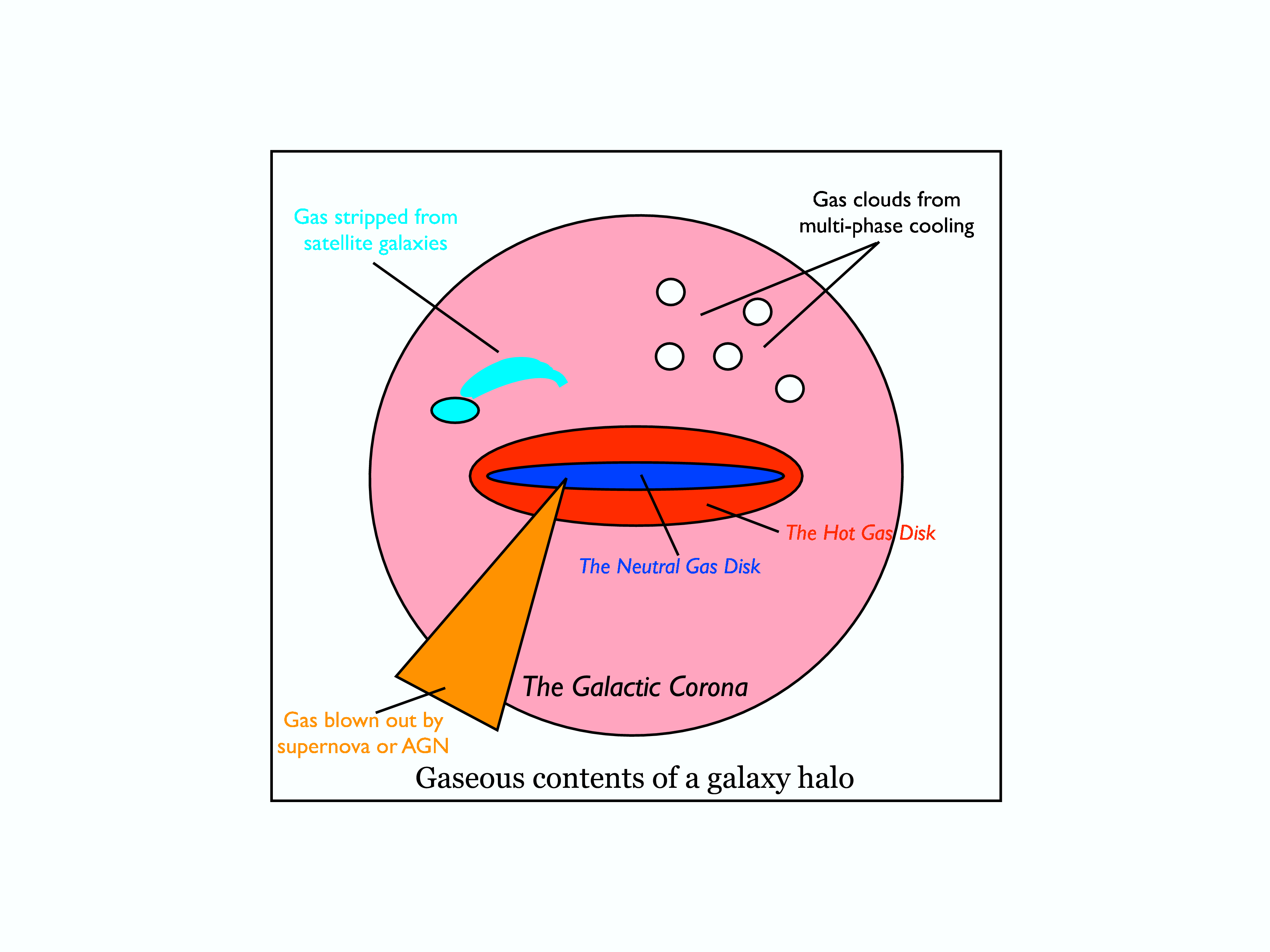}
 \vspace{1cm}
 \caption{This figure shows the possible gaseous contents of a Milky way type galaxy
halo.  The neutral gas disk and surrounding hot gas disk are both imbedded in the
galactic corona which fills the galaxy's dark matter halo.  Both the hot gas disk and the 
galactic corona have temperatures of $\sim 10^6 K$.  The galaxy's halo may also contain
gas stripped from infalling satellites, gas clouds that have cooled out of the galactic 
corona (multi-phase cooling), and gas that has been ejected from the disk because of 
energetic feedback.  All of these sources of gas may give rise to absorption systems
in quasar spectra.
}\label{fig:gh}
\end{figure}

Figure \ref{fig:gh} shows a sketch of possible components of  the gaseous content
of a galaxy's halo.  While we know a neutral gas disk is present in all spiral galaxies
locally, the hot gas disk is only detected in the Milky Way 
\cite[(Wang et al. 2005)]{wang:05}.  The existence of a galactic corona was
postulated for the Milky Way long ago \cite{spit:62}; however, it remains 
difficult to detect directly.  This same component is clearly seen in clusters and 
groups where the higher temperature makes it visible in x-rays.  Recently, there 
has been indirect evidence for the existence of a galactic corona around the
Milky Way from its interaction with high velocity clouds 
\cite[(e.g. Tripp et~al. 2003)]{tripp:03}, 

As described below, the contributions of gas stripped from satellite galaxies, cooled 
from the galactic corona or ejected by energetic feedback to the gaseous halo is still 
unclear.  While all these processes happen at some level, their relative importance is 
model dependant  and probably a function of halo mass and redshift.  Further observations 
and more detailed modeling are needed to build a more complete picture.  Below I give a 
brief description of each component and what absorption systems they maybe associated with.

\begin{itemize}

\item \textit{The Neutral Gas Disk.}  Neutral gas in a galaxy's disk is the fuel for
star formation and is observed in HI and as damped Lyman alpha and Lyman limit systems.  
However, based on the kinematics of damped Lyman alpha systems 
\cite[(Prochaska \& Wolfe 1997; 1998)]{pw:97,pw:98} it is not clear that at high redshift 
neutral gas is primarily associated with gas disks 
\cite[(Haehnelt et~al. 1998; Maller et~al. 2001)]{hsr:98,mpsp:01}.  It is also unclear 
where the transition from predominantly neutral to predominantly ionized gas occurs, 
except that it is below the defining column density of damped Lyman alpha systems.

\item \textit{The Hot Gas Disk.}  The Milky Way's disk is enveloped in a second hot 
($\sim10^6 K$) disk with a scale length $\sim1$ kpc.  The cooling times for this 
component are short, so it must be constantly refueled (or re-energized) most likely
from supernova explosions.   This component is the dominant source of x-ray absorption 
around the Milky Way making direct detection of the galactic corona and the local 
warm hot intergalactic medium extremely difficult \cite{wang:05}.

\item \textit{The Galactic Corona.} This is the relic gas from the formation of a galaxy's
dark matter halo.  In clusters and massive groups it is observed in x-ray emission.  In 
smaller mass halos cooling times are shorter and the galactic corona must have a lower
density (in the Milky Way of order $10^{-4} {\rm cm}^{-3}$).  Therefore, it is difficult to 
detect in x-rays, which is further complicated by the existence of the hot gas disk.  
In dwarf galaxies the cooling times are so short that there is effectively no hot gas 
corona.

\item \textit{Gas stripped from satellite galaxies.} Gas stripped from satellite galaxies 
and during merges will also inhabit a galaxy's halo \cite{mayer:05}. The most impressive 
example of this locally is the Magelenic Stream. This gas can be thought of as a subset
of the warm clouds infalling into the galaxy in multi-phase cooling and 
can be a source of Lyman limit and metal-line systems.

\item \textit{Gas clouds from multi-phase cooling.}  In multi-phase cooling we expect
warm clouds to condense out of the cooling galactic corona 
\cite[(Maller \& Bullock 2004; Mo \& Miralda-Escude 1996)]{mb:04,mm:96}.  These clouds
(seen perhaps as high velocity clouds in the Milky Way) may give rise to Lyman limit and
metal-line systems.  This gas can help explain the kinematics of high-ionization-state gas 
in damped Lyman alpha systems \cite{mpsp:03}. 

\item \textit{Gas ejected by supernova or other feedback.}  Supernova and other energetic
sources (e.g. AGN) can eject gas from the galaxies disk into the halo and possibly 
entirely out of the halo.  This is directly detected around star-bursting galaxies.  This 
gas can give rise to metal-line and Lyman limit systems and even possibly damped Lyman
alpha systems at high redshift.

\end{itemize}

It is important to note that low mass dark matter halos are expected to have a negligible 
mass in the galactic corona.  Therefore, there is little mass in warm clouds in these 
halos, \textit{and}  it is easier to eject gas by supernova driven winds.  Thus, it is 
likely that the content of a galaxy's halo may be a strong function of the galaxy's halo 
mass.  Most absorption systems could be the result of more than one component, complicating 
the interpretation of observations.  Therefore, it is important to look for ways 
to distinguish which component of the gaseous halo is the source of an absorption system.
Metallicities, ionization states and kinematics can all be useful diagnostics in this
endeavor.  Constraints on the properties of each component can be used to distinguish 
between models of gas cooling in galaxies and thus are central to understanding galaxy 
formation.

\section{Conclusions}

I have shown how multi-phase cooling can match the high mass end of the galaxy mass 
function above and that observations of absorption systems are an important probe of
this model.  Only from absorption systems can we probe how the warm cloud masses 
vary with halo mass and redshift.  
Furthermore, I have outlined the gaseous content of a galaxy's halo
and how absorption systems probe various components of it. Because different components
can give rise to the same type of absorption system it is necessary to use other
diagnostics to determine what component the absorption system is probing.
In summary, the study of absorption systems is crucial to understanding how gas cools 
in dark matter halos and therefore how galaxies form.

\begin{acknowledgments}
I would like to  thank the organizers of the conference for the exceptional meeting and 
the chance to visit Shanghai, China.  Also, I'd like to thank James Bullock for his 
comments on an earlier draft of this manuscript and the hospitality of the 
Center for Cosmology at the University of California, Irvine.
\end{acknowledgments}

\end{document}